# Stability of Brain Functional Network During Working Memory Using Structural Balance Theory

Sepehr Gourabi [1], Masoud Lotfalipour [1], Reza Khosrowabadi [1], Reza Jafari [1,2,3]

1 Institute for Cognitive and Brain Sciences, Shahid Beheshti University, Tehran, Iran

2 Department of Physics, Shahid Beheshti University, Evin, Tehran, Iran

3 Center for Communications Technology, London Metropolitan University, London N7 8DB, United Kingdom

## Abstract

Working memory plays a crucial role in various aspects of human life. Therefore, it has been an area of interest in different research studies, especially neuroscience. The neuroscientists investigating working memory have primarily emphasized the brain's functional modularity. At the same time, a holistic perspective is still required to investigate the brain as an integrated and unified system. We hypothesized that the brain should shift towards a more stable state during working memory than the resting state. Therefore, based on the Structural Balance Theory (SBT), we aimed to address this process. To achieve this, we examined triadic associations in signed fMRI networks in healthy individuals using the N-back as the working memory task. We demonstrated that the number of balanced triads increased during the working memory task compared to the resting state, while the opposite is true for imbalanced triads. The increase of balanced triads forced the network to a more stable state with a lower balance energy level. The increase of balanced triads was crucially related to changes in anti-synchrony to synchronous activities between the Temporal Cortex, the Prefrontal Cortex, and the Parietal Cortex, which are known to be involved in various aspects of working memory, during the working memory process. We hope these findings pave the way to a better understanding the working memory process.

**Introduction**

Working memory is an essential ability to temporarily store and manipulate information (A. Baddeley, 2000; A. D. Baddeley & Hitch, 1974). Working memory is crucial to retain and manipulate information quickly (A. Baddeley, 1996) to associate with a future action (Hooker, 1960) or be implied in complex cognitive abilities such as learning and reasoning (A. D. Baddeley & Hitch, 1974). Therefore, understanding the working memory process in the brain has been an area of interest in many studies, especially in cognitive neuroscience.

Neuroscientists have mainly focused on identifying the brain regions and their interactions in working memory using neuroimaging methods such as fMRI and brain signal recording tools like EEG. Previous fMRI studies have indicated that areas such as the Inferior frontal junctions (Jimura et al., 2018), Dorsolateral prefrontal cortex (Jimura et al., 2018; Vartanian et al., 2013), and Ventrolateral prefrontal cortex (Vartanian et al., 2013) play an essential role in working memory. These studies have linked working memory to the activities of the prefrontal cortex (Pribram et al., 1952) and confirmed it through studies using electrophysiological signals (Fuster & Alexander, 1971; Kubota & Niki, 1971). Although it has not only been limited to the frontal cortex, other brain areas have also been investigated (Brooks et al., 2020).

For instance, the impact of working memory exercises has been shown to increase fractional anisotropy, a measure of white matter density, in regions near the intraparietal sulcus and the anterior corpus callosum (Takeuchi et al., 2010). Additionally, gray matter volume, assessed through voxel-based morphometry, revealed volumetric reductions in the bilateral fronto-parietal regions and the left superior temporal gyrus (Takeuchi et al., 2011). Moreover, specific changes in white matter diffusivity have been observed in the right dorsolateral superior longitudinal fasciculus and the left Para hippocampal cingulum (Metzler-Baddeley et al., 2017).

Nonetheless, it is not only regional changes but global changes are also observed in the brain network. For instance, an increase in functional connectivity between the right medial prefrontal gyrus and other regions of the frontoparietal network, including bilateral superior frontal gyrus, paracingulate gyrus, and anterior cingulate cortex, has been reported(Jolles et al., 2013). In addition, graph theory also helped to examine this process from the global perspective, and a significant decrease in the clustering coefficient and the normalized shortest path length indicative of reduced local efficiency accompanied by increased global efficiency were reported during the N-back task, a widely used task to measure working memory performance (Yu Sun et al., 2014). Moreover, an increase in large-scale frontoparietal networks, Executive Control Networks, and Dorsal Attention Networks has also been reported (Thompson et al., 2016).

Nevertheless, the emergent properties of complex brain networks, the effects of the links on each other, and the stability of the functional brain network (Moradimanesh et al., 2021; Saberi et al., 2021a) must be well understood. The structural balance theory based on triadic associations provides an essential tool to investigate these properties in the network (Saberi et al., 2021a). Triadic associations refer to the relationships between three nodes in a network, and SBT parameters are the specific measures used to quantify these relationships. SBT suggests that a third node in a triadic structure can significantly influence the relationship between the two other nodes. The theory proposes that the relationship between two nodes is likely consistent with the triad sign. If a triad (+, +, +) or (+, -, -) exists between nodes, the probability of positive relationships between the nodes is higher. Otherwise, the probability of forming a negative relationship is more significant. In this context, the first triad type is a balanced triad, and the second type is an imbalanced triad (Belaza et al., 2017; Pham et al., 2022). In this context, when a network requires change, the number of negative links or their tendency to make a hub influence the number of balanced/imbalanced triads to force the network to another balanced energy state (Saberi et al., 2022). Such changes have been observed during emotion processing (Soleymani et al., 2023), disorders such as OCD(Talesh et al., 2023), and autism(Moradimanesh et al., 2021).

Since the brain's functional network tends to be in a more stable state during working memory (Constantinidis & Klingberg, 2016; Edin et al., 2007), we hypothesized that during working memory, the brain tends to shift towards a more stable state compared to the resting state. We aimed to investigate all the SBT parameters to indicate their specific changes during the working memory process. Therefore, SBT parameters of the brain functional network were extracted from fMRI data during working memory and were compared to the resting state parameters.

**Participants and Data Acquisition**

In order to investigate our hypothesis and implement the discussed theory, we turned to the Human Connectome Project(HCP) dataset. We utilized Functional magnetic resonance imaging (fMRI) data 3_T from 138 out of 1200 subjects. We utilized cloud-hosted data to facilitate our work due to the limited analysis capacity of our systems and the large amount of HCP data. This data, uploaded to the cloud by the Neuromatch Academy, which primarily focuses on computational neuroscience studies, was available in the Google Colab environment, providing access to the data of 339 individuals out of a total of 1200. We filtered the available data so that all our final subjects were healthy, male, and right-handed. After selecting male subjects, using the results of the Edinburgh Handedness questionnaire employed to assess handedness in this dataset, we specifically chose right-handed individuals(Oldfield, 1971). In the final stage, we examined the participants' mental health using the Mini-Mental State Examination (MMSE) results. Given that scores below 24 fall outside the normal and healthy range(Folstein et al., 1975), we excluded participants with scores below 24 from the study. Ultimately, we arrived at a cohort of 138 healthy adult subjects aged 22-35.

We continued our study with preprocessed Resting-State fMRI(rs_fMRI) and working memory data from 138 subjects sourced from the HCP dataset, focusing on extracting functional connectivity from each subject. The preprocessing procedure was carried out by HCP itself, following the defined pipeline(WU-Minn, 2017).

The rs_fMRI data had a Repetition Time (TR) of 720 ms, a Time to Echo (TE) of 33.1 ms, and a run duration of 14 minutes and 33 seconds in the study. Additionally, the working memory data, with the same TR and TE, was analyzed, with a run duration of 5 minutes and 1 second.

**Task**

According to the protocol outlined for Task-based fMRI(tfMRI) concerning the working memory task, participants were presented with blocks comprising images of places, tools, faces, and non-mutilated parts of the body (body parts without mutilation and deformities). In each run, four different types of stimuli were presented in separate blocks to individuals. Additionally, as a working memory comparison, half of each block was assigned to 2-back and the other half to 0-back. At the beginning of each block, a stimulus representing the type of task (e.g., target for 0-back) was displayed for 2.5 seconds. Each of the two runs included 8 task blocks (10 trials, each lasting 2.5 seconds for 25 seconds) and four fixation blocks (15 seconds). Each trial presented the stimulus for 2 seconds, followed by a 500-millisecond inter-task interval (ITI)(WU-Minn, 2017).

**Structural Balance Theory**

The Structural Balance Theory (SBT) examines the structural balance of systems. Fritz Heider first developed this theory within the social psychology framework, focusing on examining interpersonal relationships within a social system(Heider, 1946, 1958). Cartwright and Harary later formalized and conceptualized this theory, transforming it into a fundamental concept in network science(Cartwright & Harary, 1956).

In this theory, an examination is conducted to determine whether our targeted system is stable or unstable. According to this theory, by selecting a number of members within a network and scrutinizing the nature of relationships among them—whether they are positive links (friendships) or negative links (enmities)—we can conclude whether the entire system is stable or unstable based on these friendships and enmities(Saberi et al., 2021b).

According to the SBT, every system is classified into balanced and imbalanced. Our focus here is on triadic systems, and based on these triadic systems, we elaborate on triadic balance and imbalance. A triadic balance is a triad in which the product of the signs of its connections is positive. In contrast, an imbalanced triad is one in which the product of the signs of its connections is negative(Marvel et al., 2009).

**Types of Balanced and Imbalanced Triads**

Following the initial clarification of the SBT, theoretical physics research focused on expanding this theory. One of these studies addressed the further development of various types of balanced and imbalanced triads. Belaza and colleagues conceptualized four possible states for balanced and imbalanced triads(Belaza et al., 2017). This classification is based on the sign of connections between two triad points.

In the first type, weakly imbalanced or $T_0$, all links in the triad are negative. The second type, weakly balanced or $T_1$, occurs when one of the links in the triad is positive. The third type, strongly imbalanced or $T_2$, applies to a triad where two links are positive, and only one is negative.

The fourth type in this classification is strongly balanced, or $T_3$, where all links in the triad are positive. This categorization provides a framework for understanding different configurations of triads in terms of balance and imbalance.

**Balanced Energy**

Since we can extract the triads and examine their types using a signed network(Cartwright & Harary, 1956), we can utilize the formulation provided by Marvel and colleagues for the explanation and calculation of Balance Energy(Marvel et al., 2009). The formula for calculating Balance Energy is as follows:

$$E = -\frac{1}{\binom{N}{3}} \sum_{ijk} S_{ij} S_{ik} S_{jk}$$

Considering the provided formula for calculating Balance Energy, S represents the sign of links in the signed network, and based on its index, we can determine the sign of the desired link between which two nodes. Ultimately, to calculate the Balance Energy, we need the product of the signs of the three links in the considered triad. Then, we sum this result for all possible triads. Also, the term $\binom{N}{3}$ denotes the number of possible triads in an N-node network. The negative sign in the multiplication contributes to a better understanding of the equation from the perspective of physics about energy. It allows us to discuss and analyze the stability of the existing triad system. According to the principle of minimum energy, a physical system loses its energy and transitions from a critical state to a more stable state. Additionally, a physical system in a stable state remains constant at the minimum energy level until it receives external energy(Saberi et al., 2021a). This approach aids us in examining the level of stability in different states of the brain in our study.

**Tendency to Make Hub**

In recent years, with the development of SBT in the field of cognitive sciences, a new criterion for global hubness was introduced and named "Tendency to make hub (TMH)" to quantitatively assess the power of link aggregation around nodes in a network from a global perspective .Also, based on the sign of links, we can extend this formula for the negative links, which were crucial in this study, and formulate Negative TMH(Saberi et al., 2021a). TMH and Negative TMH for networks are defined as follows:

$$\text{TMH} = \frac{\sum_{i=1}^{N} D_i^2}{\sum_{i=1}^{N} D_i} \quad , Neg\text{TMH} = \frac{\sum_{i=1}^{N} NegD_i^2}{\sum_{i=1}^{N} NegD_i}$$

where $D_i$ represents the degree of a separate node, and N represents the total number of nodes in the network. Additionally, $NegD_i$ denotes the negative degree of that node.

**Functional Connectivity**

To investigate our hypotheses using the SBT, we needed to extract the functional connectivity matrix for each subject in two states: resting state and when individuals responded to the working memory task.

For the resting state data, time series for each participant were extracted, and using the BOLD signal in these temporal sequences, we proceeded with the analysis. Subsequently, utilizing the Glasser 360 atlas(Glasser et al., 2016) and mapping the 360-defined points in this atlas to each individual's data, we calculated Pearson correlation coefficients. These coefficients measure the strength and direction of linear relationships between different pairs of variables. In functional connectivity, higher correlation values indicate more robust functional connectivity or more remarkable similarity in the activation patterns of associated brain regions. The obtained matrix for the resting state was diagonal and symmetric.

However, since participants responded to the working memory task at the working memory data, the connectivity matrix had to be extracted based on this task. Initially, time series and temporal information related to task presentation and temporal features of presented stimuli during individuals' responses were extracted. The task and condition specified began after loading the time series data and information related to the corresponding events. Time series data record measurements or observations at specific time points, reflecting aspects of brain activity. In tandem, event information and details about experimental events, including temporal features and classifications of presented stimuli or conditions during the task, are included.

After acquiring the relevant data, the time series information was classified into predicted discrete runs on the desired frames specified and derived from the event information. This classification was performed to separate data related to individual task executions, facilitating focused analysis.

When data was divided into discrete runs, the time series data continued to be concatenated across runs and merged along the last axis. This merging yielded a combined array, `task_data,` presenting a continuous time series encompassing the entire task and specified conditions.

The final step involved calculating the connectivity matrix. This function computed Pearson correlation coefficients between different pairs of variables within the `task_data.` The resulting matrix is a connectivity image of the task between various regions or nodes at different time points. Each element within the matrix conveys the strength and direction of the correlation between corresponding pairs of variables, potentially indicating interactions in different brain measurements.

**Statistical Analysis**

In this study, we utilized various statistical methods appropriate for different sections. Initially, after extracting functional connectivity for each subject in this study, we employed the Interquartile Range (IQR) method to eliminate outliers for extracting the components of SBT. These components were necessary for investigating and elucidating structural changes in the brain during working memory. Subsequently, for each of the relevant components, we applied the nonparametric Wilcoxon signed-rank test to examine the differences in components between resting state and working memory. We used $p < 0.05$ (Fisher permutation) as a significant threshold for the observed differences. Significance levels were categorized into different groups based on the P value, and for each level, we indicated the number of stars in the boxplot chart. If p-value < 0.001, we denoted it with ***; if p-value < 0.01, with **; and if p-value < 0.05, with *.

In the next section, which aimed to identify important brain regions that played a role in working memory by changing connections from negative to positive links with other areas, we employed the paired t-test. After extracting functional connectivity for both resting state and working memory conditions, we initially examined the changes in all links in this matrix. Through this statistical test, we identified significant regions based on the brain atlas used in the study, namely Glasser 360(Glasser et al., 2016). These regions played a role in reducing the number of negative links in the resting state and transforming them into positive links during working memory.

**Result**

Our analysis observed that the number of positive links increased significantly during the working memory task in all conditions, both 0-back and 2-back, compared to the baseline resting state (p-value = $7.45 \times 10^{-13}$). On the other hand, as expected for negative links, the opposite of what happened for positive links occurred. The number of negative links showed a decreasing trend, and during individuals' responses to the task, the number of negative links was lower than the resting state (p-value = $7.45 \times 10^{-13}$).

Next, our analysis investigated the number of balanced and unbalanced triads. We found that the number of balanced triads increased during individuals' response to the working memory task compared to the resting state. This increase was statistically significant (p-value = $6.19 \times 10^{-5}$). As expected, we observed the opposite behavior or a complementary pattern for the number of unbalanced triads. The number of unbalanced triads decreased during the task compared to the resting state, and this difference was statistically significant (p-value = $6.19 \times 10^{-5}$).

Continuing our analysis, we examined the various types of balanced and unbalanced triads. After statistical analysis, we found that each of the four types of triads exhibited consistent behavior with each other. The number of $T_0$ triads compared to working memory and resting state was not statistically significant (p-value = 0.44). This is in contrast to a significant decrease for $T_1$(p-value = $1.3 \times 10^{-15}$) and $T_2$ (p-value = $5.75 \times 10^{-5}$) that we observed. This decrease in these two components is again reflected in a significant increase (p-value = $5.47 \times 10^{-11}$) in $T_3$.

This implies that the number of negative connections in the brain must have decreased, which we observed in the statistical analysis of the number of negative connections. The decrease in $T_1$ and $T_2$, followed by an increase in $T_3$, signifies that Negative TMH should exhibit consistent behavior with these changes. After statistical analysis, we found that changes in Negative TMH have significantly decreased (p-value = $5.45 \times 10^{-16}$) following the changes in the four types of balanced and unbalanced triads.

As one of the SBT parameters, Balance Energy exhibited differences between the response to the working memory task and the resting state. We observed a decrease in energy levels during the task compared to the resting state, indicating a more stable functional performance during task response, and this difference was statistically significant (p-value = $6.19 \times 10^{-5}$).

In conclusion, through a statistical examination of changes in links from resting state to working memory, we discovered that the most influential areas on other links are located in the temporal cortex, parietal cortex, and prefrontal cortex regions. In these areas, we observed that the connectivity pattern has shifted from negative to positive with other regions, leading to the emergence of stable changes in the brain through the enhancement of balanced triads.

**Discussion**

Previous studies have demonstrated that structural and functional associations among brain regions undergo significant changes during the execution of a task or cognitive process related to working memory(Engvig et al., 2010; Heinzel et al., 2017; Jolles et al., 2013; Olesen et al., 2004; Saikia, 2023). However, these studies have yet to approach these changes holistically. So far, they have yet to address the impact of the topology and arrangement of signed links (triadic associations) on the collective behavior of the brain. Therefore, in this study, we assume that the topology of signed links during individuals' response to a working memory task will exhibit a noticeable difference compared to the resting state. Additionally, due to changes in the arrangement of these links, the collective behavior of the brain in different conditions will show significant differences, ultimately leading to changes in brain stability. We employed the Structural Balance Theory (SBT) to investigate this hypothesis. SBT allows us to examine the topology of signed links, four different types of triads, and their interactions and to measure the impact of their changes on alterations in the stability or instability of the brain functional network.

**Brain Stability in Working Memory Processing**

To the best of our knowledge, the present study is the first to investigate working memory using the SBT framework in healthy individuals. Additionally, this study employs the SBT framework to analyze changes in signed links within various triads, the quantity and quality of balanced and unbalanced triads, the number of positive and negative links, and balanced energy in the brain. We aimed to demonstrate how the number of positive and negative links can influence the balance of energy in the functional brain network during working memory processing through balanced and unbalanced triads in the network. Based on previous studies, brain stability will decrease as the number of balanced triads increases concurrently with a decrease in the number of imbalanced triads (Saberi et al., 2021b). As our results indicate, in the group that has undergone the working memory task, we observed a reduction in the number of imbalanced triads and an increase in balanced triads compared to the resting state. Therefore, the brain's state during working memory-related processes appears to have greater stability than the resting-state condition. The brain maintains a more stable state when processing stimuli when performing a working memory task.

**Explanation Brain Stability**

In order to analyze how brain stability changes during the processing of working memory stimuli, our study focuses on examining the quality of triads. We chose to investigate the quality of triads because the change in brain stability during the resting state to a more stable state in the group that performed the task is related to alterations in the quantity of balanced and imbalanced triads. Based on previous studies, we know that the quality of triadic associations depends on negative and positive links and their topology(Moradimanesh et al., 2021; Saberi et al., 2021b). Therefore, the number of imbalanced triads, namely $T_0$ and $T_2$ (imbalanced triads) and $T_1$ and $T_3$(balanced triads) in the resting-state and task-performing groups, is compared, as described in the methods section.

As explained, the number of imbalanced triads in the task-performing group significantly decreased compared to the resting state, and there was an increase in the number of balanced triads in this group. The analysis of results indicates that this decrease in the number of imbalanced triads and the increase in balanced triads are due to changes in triads $T_1$, $T_2$, and $T_3$. These results suggest that triads $T_1$ and $T_2$ in the resting state have transformed into triads of type $T_3$ in the task state. Therefore, these changes in balanced and imbalanced triads lead to the emergence of a stable brain state during the task compared to the resting state. From this perspective, these changes resulting in brain stability during the task have led to the processing of working memory stimuli. In other words, changes in the number of triadic associations have induced a demand for a change in brain state.

**The Role of Negative Tendency to Make Hub (TMH)**

Since one of the critical findings of our study is a reduction in the number of imbalanced triads and an increase in balanced triads as a result of changing the sign of links within the functional brain network, ultimately leading to enhanced brain stability, One of the hypotheses of our study is the tendency of the links to form a hub in order for the brain to achieve this stable state. We utilized a global hubness measure called "tendency to make hub" (TMH) to investigate this hypothesis. TMH serves as a metric to assess the strength of link gathering around nodes in a network from a global perspective(Saberi et al., 2021b).

Based on the analysis of the results of our study, we observed a decrease in negative TMH within the functional brain network. This reduction in negative TMH aligns with our other study findings, specifically the increase in balanced triads. The simultaneous increase in balanced triads and decrease in negative TMH both contribute to an overall increase in brain stability. Therefore, the reduction in negative TMH, coupled with the increase in balanced triads, promotes brain stability. Accordingly, as quantified by TMH, links' negative tendency to form hubs is associated with the drive toward increased brain stability. The interplay between these measures highlights the intricate dynamics of network organization and its impact on the stability of the functional brain network.

**Role of Cortical Regions in Changes to the Stability of the Brain's Functional Network**

From our study's perspective, negative and positive links in the brain indicate neural activity. As our study aimed to investigate the role of negative and positive links in the stability of the brain's functional network, we endeavored to identify the cortical brain regions involved in this stability. Considering the results of our study, as discussed earlier, where the number of balanced triads of type $T_3$ increased, signifying changes in negative links within triads of types $T_2$ and $T_1$ and their transformation into positive links, it seems plausible that these changes could reveal the neural underpinnings of such alterations. Our analysis revealed that the most significant changes from negative to positive links occurred in the temporal regions of the brain's cortex. This observation aligns with other studies utilizing different modalities, such as Task-Based fMRI Studies(Miró-Padilla et al., 2019; Schweizer et al., 2013)and Structural Imaging Studies(Takeuchi et al., 2011), that have also pointed to the involvement of temporal regions. Additionally, parietal regions play a substantial role in these changes, consistent with prior studies(Olesen et al., 2004; Sayala et al., 2006). For instance, Constantinidis and Klingberg state that parietal cortices play a crucial role in working memory as part of a network associated with working memory, known as the fronto-parietal network. The authors suggest that more robust functional connectivity between the prefrontal cortex (PFC) and parietal cortex leads to a higher firing rate in parietal networks, resulting in increased stability in networks encoding information stored in working memory(Constantinidis & Klingberg, 2016). These findings align with our results, indicating increased $T_3$ triads in parietal regions and enhanced network stability. Furthermore, our analysis highlights the undeniable role of prefrontal cortical regions, alongside temporal and parietal regions, in mediating changes from negative to positive links. This observation is consistent with previous fMRI studies(Hempel et al., 2004; Schweizer et al., 2013), supporting the significant role of prefrontal regions in our findings related to the stability of the brain's functional network due to the increased occurrence of $T_3$ triads in parietal regions.

In conclusion, it appears that regions of the brain serve as the neural substrate for changes from negative to positive links, ultimately stabilizing the brain's functional network, including the temporal, parietal, and prefrontal cortices.

**Data Availability**

The data and analysis code used in this study are available upon request. Interested individuals can contact the author via email at sgourabi@binghamton.edu .

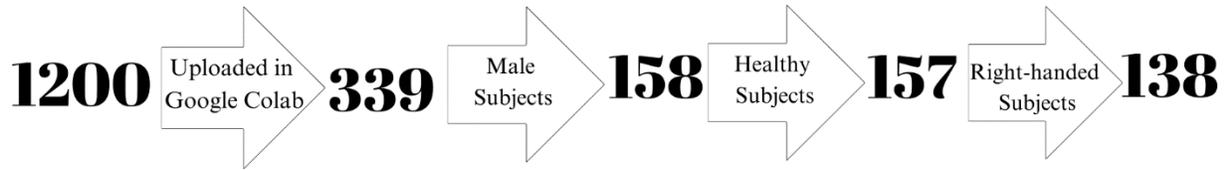

*Figure 1 _ Participant selection process*

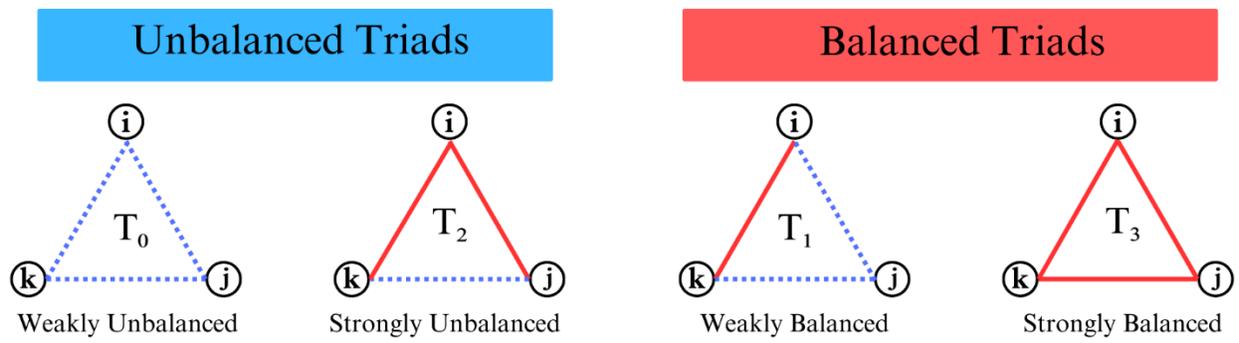

*Figure 2 _ Types of Balanced and Imbalanced Triads*

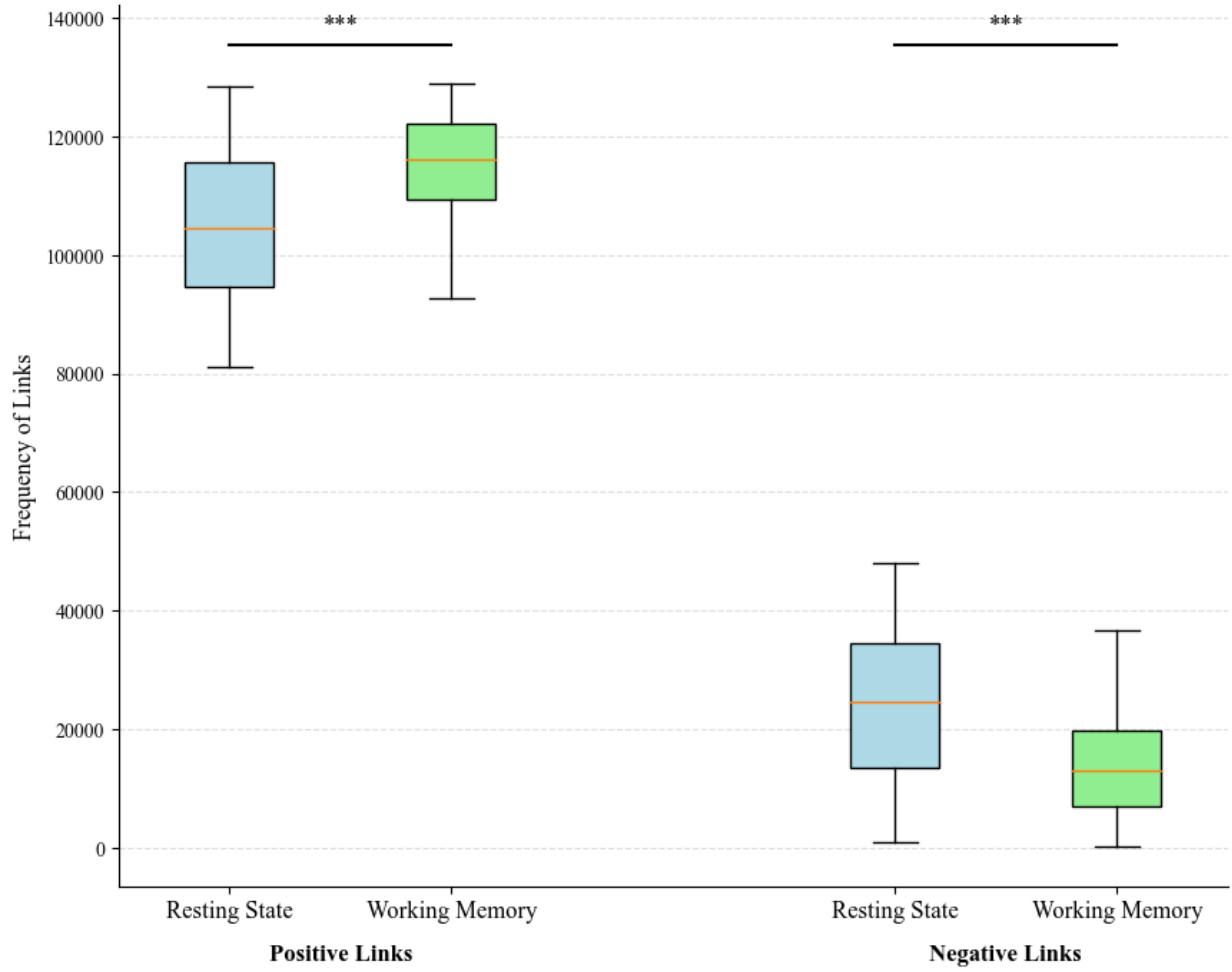

*Figure 3 _ Box Plot of Positive and Negative Links during Resting State and Working Memory*

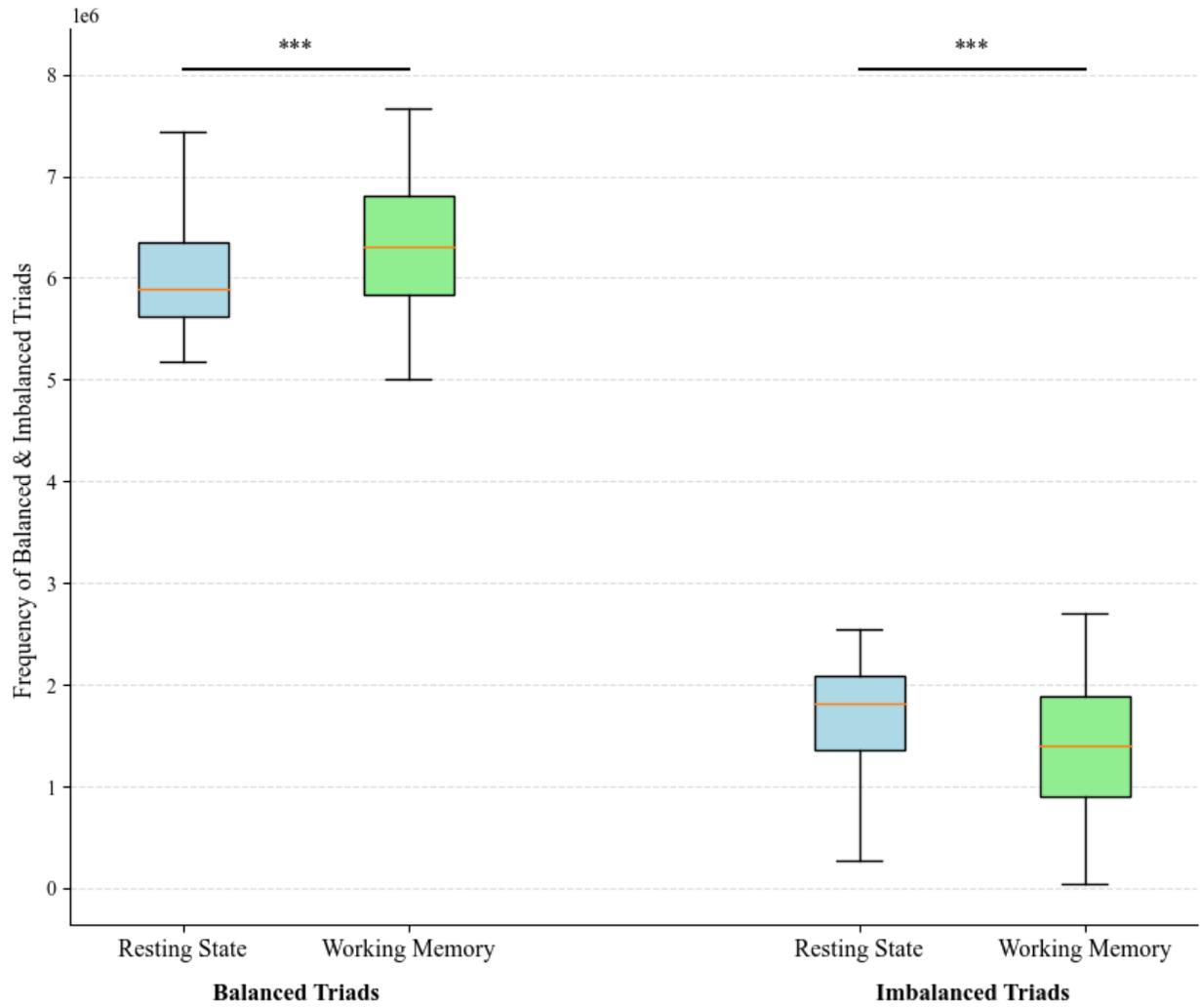

*Figure 4 _ Box Plot of Balanced & Imbalanced Triads during Resting State and Working Memory*

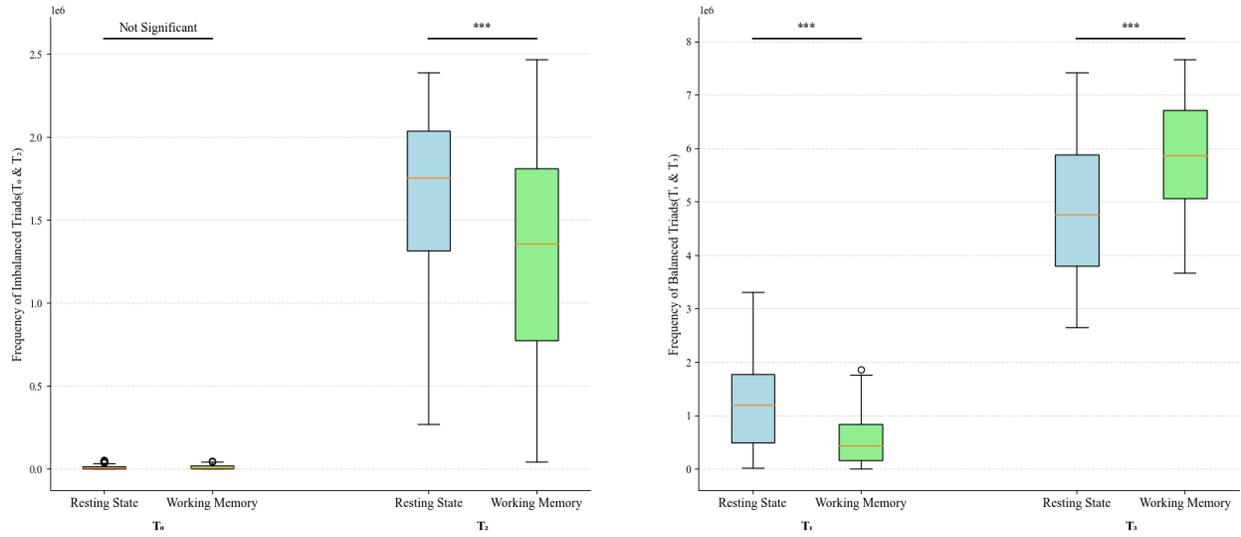

*Figure 5 _ Box Plot of Types Triads during Resting State and Working Memory*

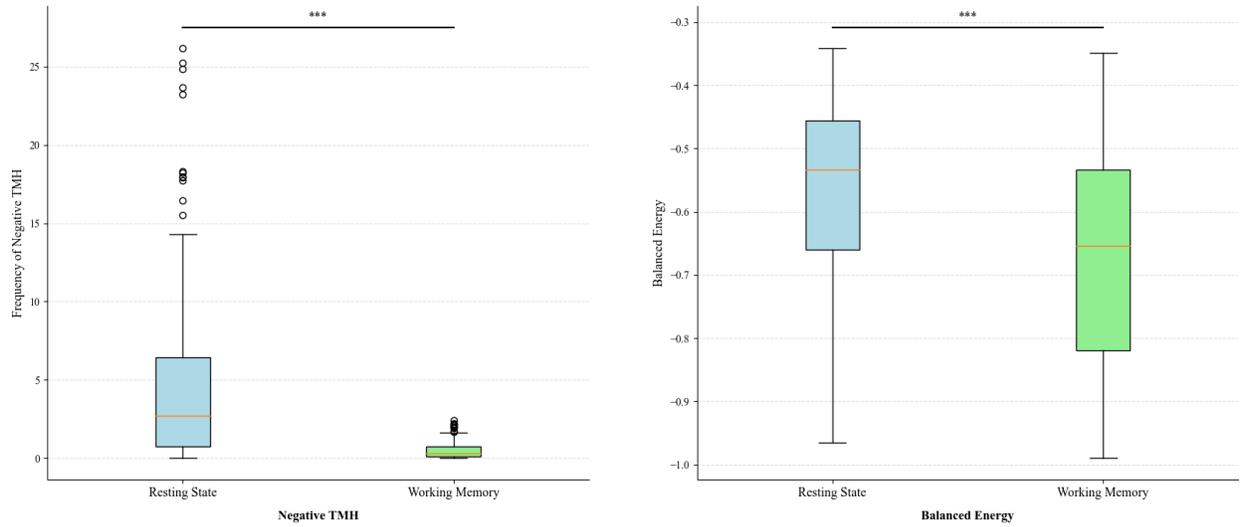

*Figure 6 _ Box Plot of Negative TMH & Balanced Energy during Resting State and Working Memory*

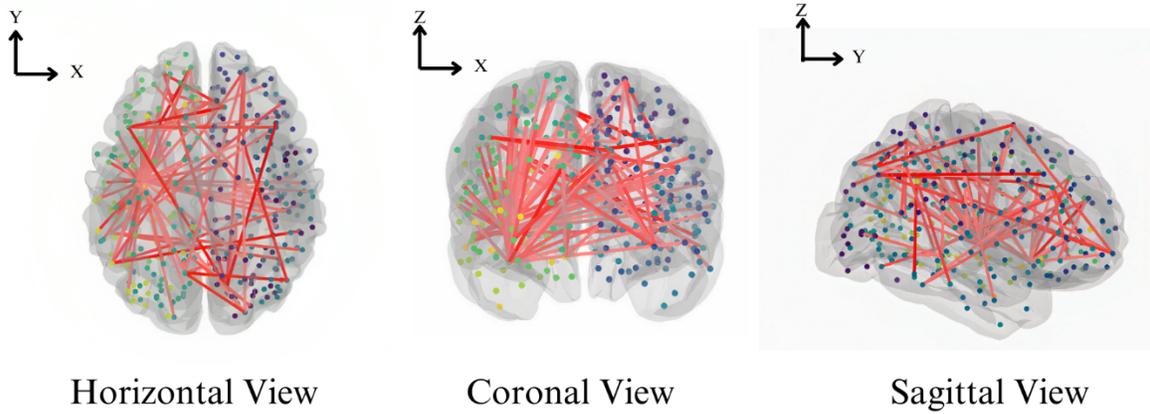

*Figure 7 _ Influential Regions During Working Memory*

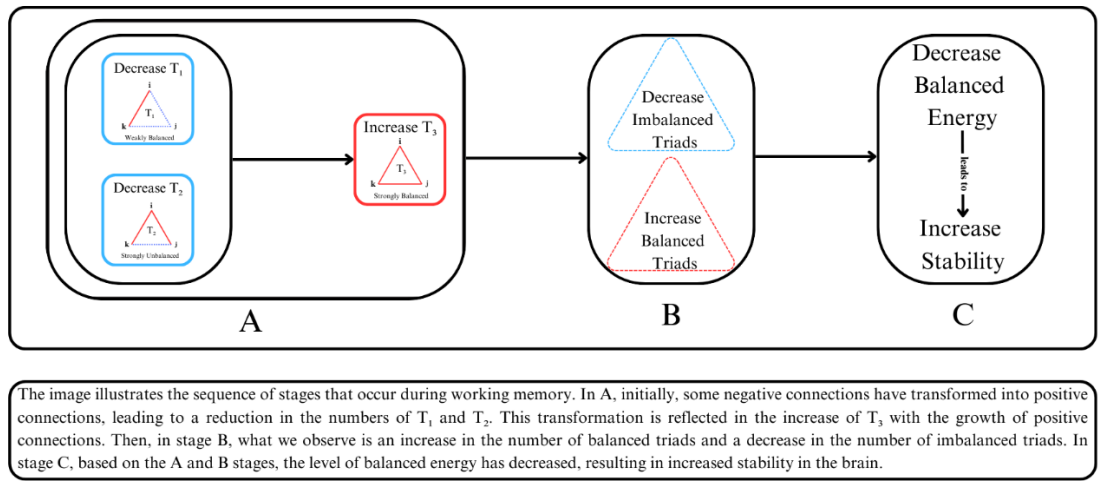

*Figure 8 _ Visualization of sequence of Stages During Working Memory*